# Fractal photon sieve

**Fernando Giménez**

*Departamento de Matemática Aplicada, Universidad Politécnica de Valencia, E-46022 Valencia, Spain*

**Juan A. Monsoriu**

*Departamento de Física Aplicada, Universidad Politécnica de Valencia, E-46022 Valencia, Spain*

**Walter D. Furlan and Amparo Pons**

*Departamento de Òptica, Universitat de València, E-46100 Burjassot, Spain*
walter.furlan@uv.es

**Abstract:** A novel focusing structure with fractal properties is presented. It is a photon sieve in which the pinholes are appropriately distributed over the zones of a fractal zone plate. The focusing properties of the fractal photon sieve are analyzed. The good performance of our proposal is demonstrated experimentally with a series of images obtained under white light illumination. It is shown that compared with a conventional photon sieve, the fractal photon sieve exhibits an extended depth of field and a reduced chromatic aberration.

**References and Links**

1. S. Wang and X. Zhang, "Terahertz tomographic imaging with a Fresnel lens," Opt. Photon. News **13**, 59 (2002).
2. Y Wang, W. Yun, and C. Jacobsen, "Achromatic Fresnel optics for wideband extreme-ultraviolet and X-ray imaging," Nature **424**, 50-53 (2003).
3. L. Kipp, M. Skibowski, R.L. Johnson, R. Berndt, R. Adelung, S. Harm, and R. Seemann, "Sharper images by focusing soft x-rays with photon sieves," Nature **414**, 184-188 (2001).
4. R. Hyde, "Eyeglass. 1. Very large aperture diffractive telescopes," Appl. Opt. **38**, 4198-4212 (1999).
5. G. Andersen, "Large optical photon sieve," Opt. Lett. **30**, 2976-2978 (2005).
6. R. Menon, D. Gil, G. Barbastathis, and H. Smith, "Photon sieve lithography," J. Opt. Soc. Am. A 22, 342-345 (2005).
7. G. Saavedra, W.D. Furlan, and J.A. Monsoriu, "Fractal zone plates," Opt. Lett. **28**, 971-973 (2003).
8. W.D. Furlan, G. Saavedra, and J.A. Monsoriu, "Fractal zone plates produce axial irradiance with fractal profile," Opt.& Photon. News **14**, 31 (2003).
9. J.A. Monsoriu, G. Saavedra, and W.D. Furlan, "Fractal zone plates with variable lacunarity," Opt. Express **12**, 4227-4234 (2004) http://www.opticsinfobase.org/abstract.cfm?URI=oe-12-18-4227.
10. W.D. Furlan, G. Saavedra and J.A. Monsoriu are preparing a paper to be called "Imaging with fractal zone plates."
11. Q. Cao and J. Jahns, "Focusing analysis of the pinhole photon sieve: individual far field model," J. Opt. Soc. Am. A **19**, 2387-2393 (2002).
12. Q. Cao and J. Jahns, "Non paraxial model for the focusing of high-numerical-aperture- photon sieves," J. Opt. Soc. Am. A **20**, 1005-1012 (2003).
13. The spectral sensitivity of the camera used in the experiment can be looked up at the site http://astrosurf.com/buil/350d/350d.htm.
14. M.J. Simpson and A.G. Michette, "Imaging properties of modified Fresnel zone plates," Opt. Acta **31**, 403-413 (1984).

## 1. Introduction

A renewed interest in diffractive focusing elements has been experienced by the scientific community in the last years because these elements are essential in image forming setups that are used in THz tomography [1], soft X-ray microscopy [2,3], astronomy [4,5], and lithography [6]. Following this trend, our group has recently introduced a new type of 2D photonic-image-forming structures: the Fractal Zone Plates (FZPs) [7-9]. When illuminated by a plane wavefront, a FZP produces multiple foci along the optical axis. The internal structure of each focus exhibits a characteristic fractal structure reproducing the self-similarity of the originating FZP. We have shown that the number of foci and their relative amplitude can be modified with the FZP design [9]. It has been suggested, and proved very recently [10], that this property can be profited in image forming systems to obtain an enhancement of the depth of field.

Photon sieves [3] are another new kind of diffractive optical elements, developed for focusing and imaging soft X rays with high resolution capabilities. A conventional photon sieve is essentially a Fresnel zone plate where the clear zones are replaced by great number of non overlapping holes of different sizes. Following the initial paper in *Nature*, several theoretical and experimental works related to photon sieves have been performed, explaining their properties from different points of view [11,12] and showing their good performance in different applications [5,6].

In this paper a new diffractive element: the fractal photon sieve (FPS) is presented. It consists of hundreds of small circular holes distributed over the zones of a FZP. We show that FPS potentially improves the performance of FZP in several aspects. Two features that greatly reduce the fabrication constraints and allow fractal focusing of electromagnetic waves in a wide range of the spectrum are the following: 1) The hole diameter can be bigger than the width of the underlying zone, allowing a better resolution as compared with a FZP made with the same technology. 2) A FPS has no connected regions and then, it can be fabricated in a single surface without any substrate. In addition, the number and the distribution of holes per zone can be modified to improve the suppression of secondary maxima and higher orders of diffraction.

## 2. Fractal photon sieve design

As was discussed in Ref. [7], a FZP can be constructed following the same procedure performed to design conventional Fresnel zone plates. Let us review the concept: As it is well known, a Fresnel zone plate consists of alternately transparent and opaque zones whose radii are proportional to the square root of the natural numbers, thus it can be generated from a 1-D structure (see Fig. 1, upper part) defined by the periodic function $q(\varsigma)$, by performing a change of coordinates $\varsigma=(r_0/a)^2$ and by rotating the transformed 1-D function around one of its extremes. The result is a Fresnel zone plate having a radial coordinate $r_o$ and an outermost ring of radius $a$ [see Fig. 2(a)]. In a similar way a FZP is constructed by replacing the periodic function used in the generation of a Fresnel zone plate, by a 1-D fractal structure, as for example the triadic Cantor set shown in Fig. 1 (lower part). The corresponding zone plate with fractal profile is represented in Fig. 2(b). It has been shown [7,10] that the irradiance along the optical axis produced by a FZP presents multiple foci with a distinctive fractal structure. The position, size and depth of the foci depend on the fractal level and on the lacunarity of the encoded fractal structure. Based on a FZP, the FPS here proposed combines the features of FZP with the concept of photon sieve. A FPS has essentially the same structure of a FZP but instead of transparent rings the corresponding zones have been broken up into isolated circular holes. The result is shown in Fig. 2(c). In the construction procedure we adopted the results reported in Ref [3] where it has been shown that for a photon sieve constructed with a Fresnel zone plate structure, the diameter $d$ of the holes in each ring of width $w$ of has an optimum value for the effective contribution to the focus. This value is given by $d=1.53w$.

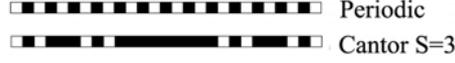

Fig. 1. Periodic and fractal 1-D structures to be used to generate zone plates.

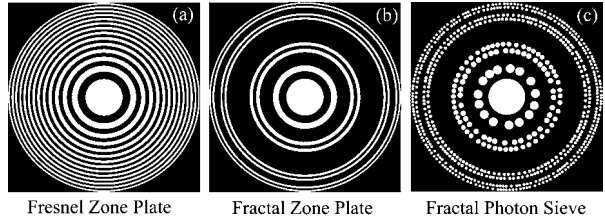

Fig. 2. Comparison between (a) Fresnel zone plate, (b) FZP, and (c) FPS.

As in a conventional photon sieve, the angular distribution of the holes in each ring can be fixed, or random [like in Fig. 2(c)]. We will show in the next section that this election is irrelevant for the axial response.

### 3. Focusing and imaging properties of FPS

Let us consider the irradiance along the optical axis, $z$, given by an optical system having a 2-D pupil function $p(r_o, \phi)$, expressed in canonical polar coordinates, when it is illuminated by a plane wave of wavelength $\lambda$:

$$I(z) = \left(\frac{2\pi}{\lambda z}\right)^2 \left| \int_0^a \int_0^{2\pi} p(r_o, \phi) \exp\left(-i\frac{\pi}{\lambda z} r_o^2\right) r_o \, dr_o \, d\phi \right|^2 \tag{1}$$

The axial irradiance in Eq. (1) can be conveniently expressed in terms of a single radial integral, by performing first the azimuthal average of the pupil function $p(r_o, \phi)$:

$$p_o(r_o) = \frac{1}{2\pi} \int_0^a p(r_o, \phi) \, d\phi . \tag{2}$$

Then Eq. (1) can be rewritten as

$$I(z) = \left(\frac{2\pi}{\lambda z}\right)^2 \left| \int_0^a p_o(r_o) \exp\left(-i\frac{\pi}{\lambda z} r_o^2\right) r_o \, dr_o \right|^2 . \tag{3}$$

It is important to note that the azimuthal average of the pupil is the final responsible of the behavior of the axial irradiance, instead of the pupil itself. Therefore, from this point of view a photon sieve can be designed to achieve any convenient apodization by a proper modulation of the density of pinholes in each zone in which the distribution of holes can be either regular or random.

To compare the performance of a FPS with the conventional FZP [7] we used Eqs. (2) and (3) to compute the axial irradiances provided by the corresponding pupil functions [see Figs. 2(b) and 2(c)]. The result is shown in Fig. 3. The parameters used to calculate these plots were $a$=2mm and $\lambda$=632.8nm. The number of holes in the FPS was 650 and the density of holes per zone (i.e. the ratio between the area covered by the holes and the total area of the zone) was aproximately 90%. The minimun diameter in the outermost ring was $d$=1.53$w$=0.0572mm. As shown in Ref. [7] the focal structure of a FZP along the optical axis is characterized by the coincidence of the central peak with the one obtained for a conventional Fresnel zone plate, but the internal structure of the focus reproduce the self-similarity of the zone plate.

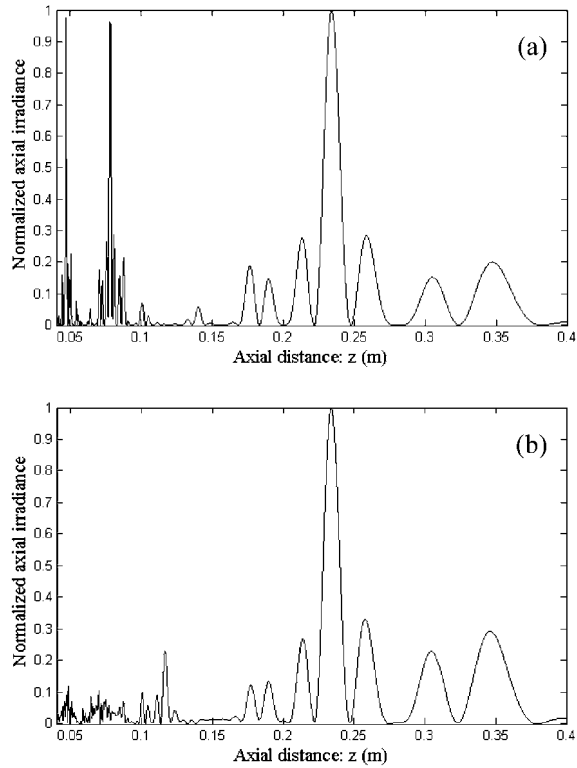

Fig. 3. Normalized irradiance distributions along the optical axis produced by the zone plates in Figs. 2(b) and 2(c). (a) FZP, (b) FPS (both computed for S=3).

This behavior is repeated at the higher-order diffraction foci [see Fig. 3(a)]. When the FZP is replaced by a FPS the principal focus remains almost unaltered but all odd higher orders are highly reduced due to the smoothing effect that the holes produce on the azimuthal average of the effective pupil [see Fig. 3(b)]. This effect is obtained at the expense of the appearance of low intensity even orders, which by design are null in the case of azimuthally uniform FZP. Note also that the secondary maxima in the principal focus are relatively higher for a FPS. We have experimentally tested the imaging capabilities of FPS under white light illumination. For comparison images of test object (consisting of binary letters from an optotype-like chart) were formed both with a conventional Fresnel photon sieve [3,5] with 81 zones and with the equivalent FPS (i.e. the same diameter and same outermost pinhole diameters) constructed for S=4. The diameter of the binary zone plates used in the experiment were 5mm and their focal distances 124mm, for $\lambda$=632.8nm. The photon sieves were printed and then photographically reduced onto 35mm slides. The images of the test object were obtained directly onto an 8 megapixels CMOS detector of area 22.2×14.8mm$^2$ (Cannon EOS 350D digital camera [13]). The results are summarized in Fig. 4. Due to the different transmittances of both kind of zone plates the range of intensities of the photographs in this figure were normalized to the peak intensity, but no additional post-processing was performed. As can be seen the out of focus image obtained with the FPS is considerably better than the one obtained with the Fresnel photon sieve (noticeable in the second and third lines of letters), the price paid to gain depth of field is a slightly poor resolution at the in-focus plane [see Figs. 4(a) and 4(b)]. Note also that the gain of the depth of field results in a considerable reduction in chromatic aberration when using white light. In spite of being a subjective comparison the result is consistent with the prediction that can be done with the hypothesis of theoretical irradiances computed using Eq. (3). This result is shown in Fig. 5 were, for simplicity, the plots have been computed for S=3

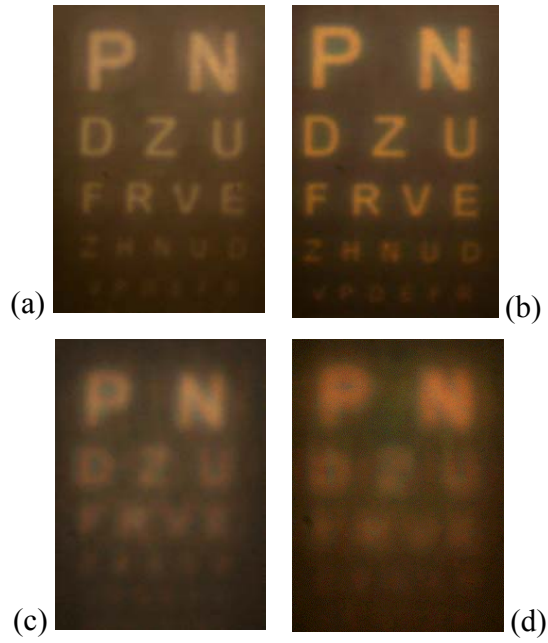

Fig. 4. Images obtained with FPS (left) and with Fresnel photon sieve (right) in a 4 $f$ setup ($\lambda$=568nm). Images (a) and (b) were obtained at the green image plane. In (c) and (d) the object-image distance was kept constant and the defocus was obtained by moving the photon sieve 45 mm towards the CCD.

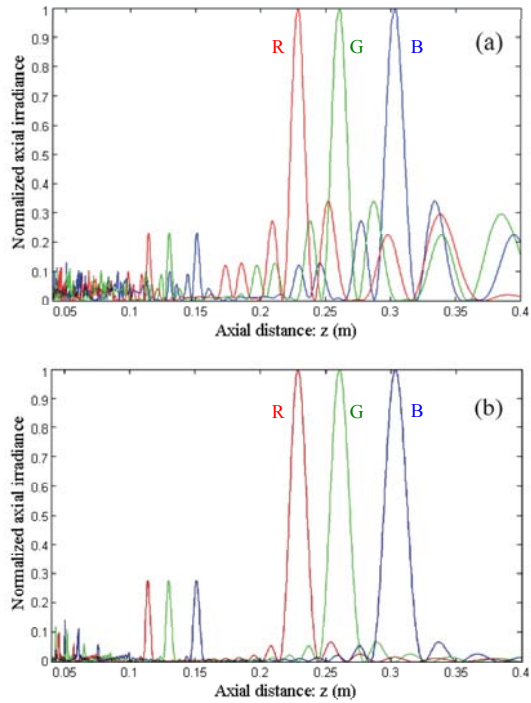

Fig. 5. Axial irradiances computed for (a) the Fresnel photon sieve and (b) the FPS used in Fig. 4 for $\lambda$=647nm (red line, R), 568nm (green line, G), and 488nm (blue line, B).

As can be seen in Fig. 5 the reduction of the chromatic blur can be explained by the overlapping of the foci for the different wavelengths that creates an overall extended depth of field which is less sensitive to the chromatic aberration.

## 4. Conclusions

We have proposed a photon sieve with fractal focusing properties. The structure of the sieve is based on FZP, and therefore, their behavior for the first diffraction order is similar. The main difference is that the higher-orders obtained with the FZP are highly reduced with the sieve. The numerical and experimental results provided in this paper show the focusing and image-forming properties of FPS. With a FPS a substantial increase in the depth of field and a noticeable reduction in the chromatic aberration can be obtained with respect to a Fresnel photon sieve of the same focal distance. In our opinion the FPS offer a great versatility in design pupils for particular focusing properties because of there are many parameters, as the fractal dimension, the number of zones, the diameter and density of holes per zone, that can be varied conveniently to obtain a specific result. Another advantage of FPS over the FZP arises from the fabrication point of view: the FPS can be constructed in a single structure without any supporting substrate. Furthermore, there are potential improvements in the design of FPS such as increasing the efficiency by the use of composite zones [14]. Therefore applications in optical and non-optical wavelength range in which graded amplitude pupils are difficult or even impossible to construct will greatly will benefit from these new focusing structures.

**Acknowledgements**

This research has been supported by the following grants:
- DPI 2003-04698, Plan Nacional I+D+I, Ministerio de Ciencia y Tecnología. Spain.
- Programa de Incentivo a la Investigación de la UPV 2005, Vicerrectorado de Innovación y Desarrollo, Universidad Politécnica de Valencia, Spain.
- MTM 2004-06015-C02-01, DGI (Spain) and FEDER Project.